\documentclass{pasj01} 
\begin{document}
\title{{\small Letters}\\
Highest-Resolution Rotation Curve of the Inner Milky Way
proving the Galactic Shock Wave}
\author{Yoshiaki \textsc{Sofue}$^{1}$ }  
\altaffiltext{1}{Institute of Astronomy, The University of Tokyo, 2-21-1 Mitaka, Tokyo 181-8588, Japan } 
\email{sofue@ioa.s.u-tokyo.ac.jp}
\maketitle 
\def\Msun{M_\odot} 
\def\deg{^\circ}  \def\Vrot{V_{\rm rot}}  \def\vrot{\Vrot}  
\def\CO{$^{12}$CO$(J=1-0)$ }     \def\Tb{T_{\rm B}}
 \def\Htwo{H$_2$ }  \def\Msun{M_\odot}\def\deg{^\circ}  \def\deg{^\circ} 
 \def\co{$^{12}$CO  }  \def\coth{$^{13}$CO  } \def\coei{C$^{18}$O }   \def\cotw{$^{12}$CO ($J$=1-0) } 
 \def\kms{km s$^{-1}$}   \def\Ico{I_{\rm CO}} \def\Msun{M_\odot} \def\msun{M_\odot} 
\def\Vsun{ V_0 } \def\Rsun{ R_0 } \def\rzero{R_0} \def\vzero{ V_0 }     
 \def\cos{~{\rm cos}~} \def\sin{~{\rm sin}~}  
 \def\Tb{T_{\rm B}}  
 \def\Vlsr{v_{\rm LSR}} \def\vlsr{\Vlsr}
 \def\vs{\vskip -2mm}
 \def\red{\textcolor{red}}
 \def\cyan{\textcolor{cyan}}
 \def\red{ }
 \def\cyan{ }
   
\begin{abstract} We present a rotation curve (RC) of the inner Galaxy of the 1st quadrant at $10\deg \le l \le 50\deg ~ (R=1.3-6.2~{\rm kpc})$ with the highest spatial (2 pc) and velocity (1.3 \kms) resolutions.
We used the \cotw-line survey data observed with the Nobeyama 45-m telescope at an effective angular resolution of $20''$ (originally $15'')$, and applied the tangent-velocity method to the longitude-velocity diagrams by employing the Gaussian deconvolution of the individual CO-line profiles.
A number of RC bumps, or local variation of rotation velocity, with velocity amplitudes $\pm \sim 9$ \kms and radial scale length $\sim 0.5-1$ kpc are superposed on the mean rotation velocity.
The prominent velocity bump and corresponding density variation around $R\sim 4$ kpc in the tangential direction of the Scutum arm (4-kpc molecular arm) is naturally explained by an ordinary galactic shock wave in a spiral arm with small pitch angle, not necessarily requiring a bar-induced strong shock.
Tables of RC are available at the PASJ supplementary data site. 
\end{abstract}
 

\section{Introduction} 

Various methods to derive the rotation curve (RC), or circular velocities in the Galactic disk, of the Milky Way have been proposed such as the tangent-velocity 
method (TVM) for gaseous disk, 
radial-velocity plus distance method for stars, 
trigonometric method for maser sources, 
and disk-thickness method for HI disk 
(see reviews by \cite{Fich+1991,SofueRubin2001,Sofue2017,sofue2020}). 
Large-scale compilation of RC data has been obtained and available 
electronically \citep{huang+2016,Iocco+2015,Pato+2017a,Pato+2017b,Krelowski+2018,sofue2020}.

The TVM measures the terminal velocity of the gaseous disk of the inner Milky Way inside the Solar circle in the HI and CO line emissions
\citep{Burton+1978,Clemens1985,alvarez+1990,mcclure+2007,marasco+2017}.
It has the advantage to uniquely determine the galacto-centric distance by 
$R=\rzero \sin ~ l$
without suffering from uncertainty in distance measurements, where $\rzero$ is the Solar circle radius.
The rotation velocity is given by 
$\Vrot=\vlsr+\vzero \sin~l,$
where $\vzero$ is the Sun's circular velocity and $\vlsr$ is the radial velocity of the object with respect to the local standard of rest.
In this paper we adopt the galactic constants of $(\rzero,\vzero)$ = (8.0 kpc, 238 \kms) \citep{Honma+2012,Honma+2015}.

Further advantage to use the CO-line is that it measures the motion of molecular clouds, which are the most massive and individual objects sharply concentrated near the Galactic plane, having the lowest velocity dispersion among Galactic objects.
Namely, CO traces the rotational kinematics of the Galactic disk most precisely with the minimum influence by random motion and velocity dispersion as possessed by other species.

In this paper we apply the TVM to the CO-line data of the inner Galaxy obtained by the FUGIN (Four-beam receiver system Unbiased Galactic-plane Imaging survey with Nobeyama 45-m telescope) \citep{mina+2016,ume+2017}.
We aim first at providing with an RC at the highest spatial (2 pc) and velocity (1.3 \kms) resolutions of the inner Milky Way in the 1st quadrant at $10\deg \le l \le 50\deg$, or at galacto-centric distance of $R=1.3$ to 6.1 kpc.
It covers the tangential directions of the 4-kpc (Scutum, $l\sim 30\deg$) arm nesting the star forming complex W43 and of the 3-kpc expanding (Norma,$\sim 20\deg$) arm.
We then discuss rotational fluctuations in relation to the kinematics of the spiral arms and galactic shock (GS) waves. 

\section{Tangent-Velocity Method to Determine Circular Velocities}

\subsection{Data}
The FUGIN survey covered the Galactic disk in the first quadrant at $l=10\deg \le 50\deg$ and $|b|\le 1\deg$.
The \cotw line channel maps had a grid spacing of $8''.5\times 8''.5\times 0.65 ~{\rm km ~s^{-1}}$ in the $(l,b,\vlsr)$ space.
The effective velocity resolution was 1.3 \kms, the rms noise of the brightness temperature $\Tb$  was $\sim 1$ K, and the effective angular resolution was $20''$, while the original beam of the 45-m telescope at the \cotw frequency was $15''$.
Although FUGIN data include \coth and \coei lines, we here analyze the \cotw line data alone, because the kinematics of the Galactic molecular disk may not be dependent on the C and O isotopes.
Moreover, \coth and \coei lines represent higher-density cores, and hence they pick up more patchy structures than those mapped by the \cotw line.

\subsection{Terminal velocities in longitude-velocity (LV) diagram} 

\begin{figure}  
    \begin{center}
    \includegraphics[width=7cm]{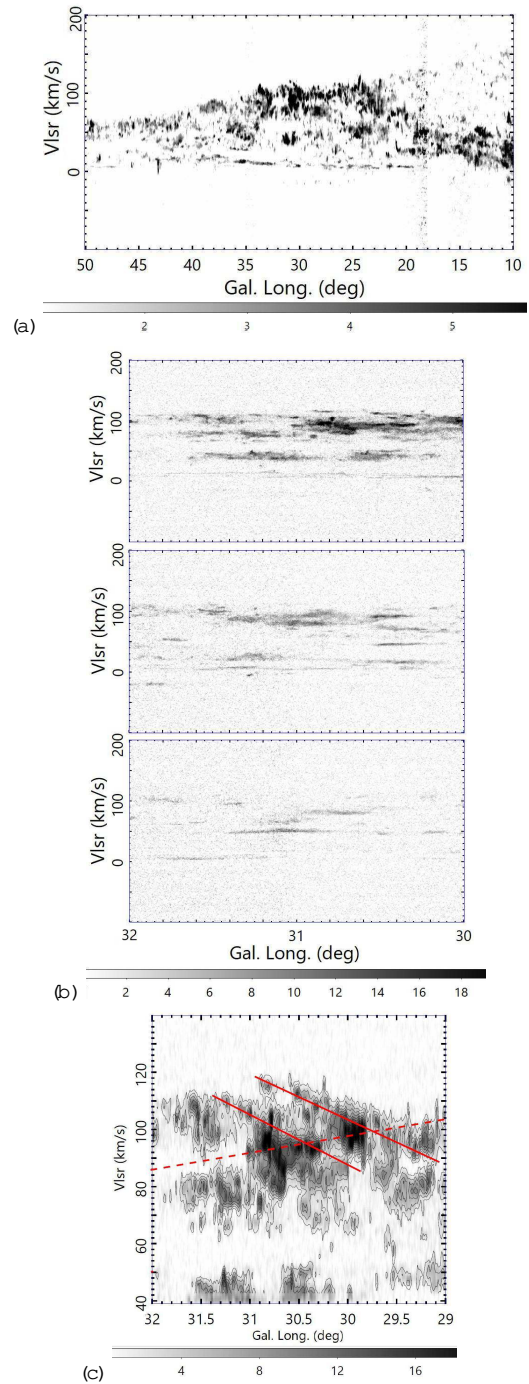}  
\end{center} 
\caption{(a) CO-line LV diagrams $l=10\deg$ to $50\deg$ at $b=0\deg$, smoothed in the longitude by Gaussian beam with $1'.4$ (10 pixels). Note that the $l$ smoothing is not applied in the analysis for RC.  
(b) LV diagrams from $l=30\deg$ to $32\deg$ at three different latitudes, $b=0\deg$ (top),  $+0\deg.5$ (middle), and $+0\deg.75$. (c) Enlargement around W43 in the 4-kpc molecular arm. The terminal edge is inclined by $\sim +10$ \kms per degree (red lines) in the opposite sense to that for flat rotation with $-3.6$ \kms per degree  (dashed line).
} 
\label{fig1} 
\end{figure}

Tangent velocities are obtained by tracing the upper-most edges of the emission regions in longitude-velocity diagrams.
Figure \ref{fig1}(a) shows LV diagram (brightness temperature $\Tb$ of the \cotw line emission against longitude) in the galactic plane, and (b) shows those around $l=31\deg$ at different latitudes, $b=0\deg, 0\deg.5$ and $0\deg.75$.
The apparent upper bounds of the LV ridges, or the terminal velocities, seem to systematically decrease with the latitude, which will be measured more quantitatively in subsection \ref{lvplot}.  

Figure \ref{fig1}c shows an LV diagram across the star forming complex W43 in the tangential direction of the 4-kpc arm. 
The intensity distribution is clumpy due to giant molecular clouds and the terminal velocity is locally variable with longitude, exhibiting LV ridges with positive gradients as indicated by the red lines, opposite to the general decrease with longitude shown by the dashed line.
The local LV variation will be discussed in detail in subsection \ref{spiralarm} in relation to the spiral arm and galactic shock wave. 

\subsection{Gaussian deconvolution of line profiles}

\begin{figure} 
\begin{center}    \includegraphics[width=8cm]{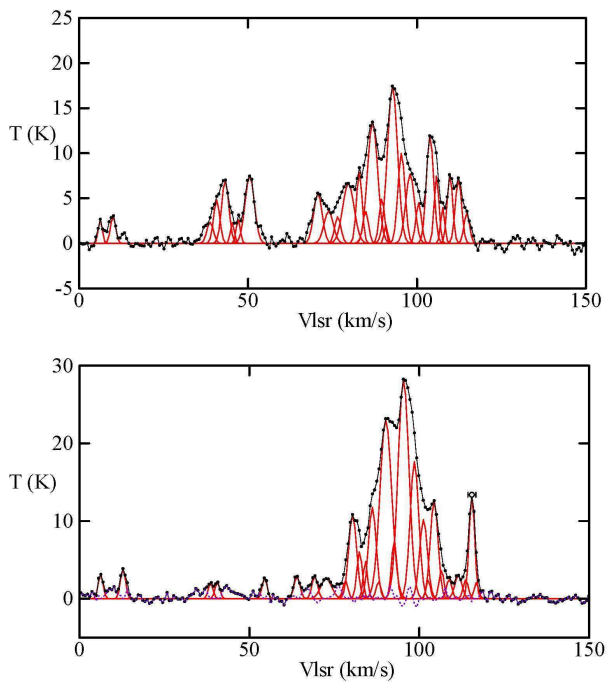}  
\end{center} 
\vs \caption{Typical FUGIN \cotw-line spectra at   $l=30\deg.4$ and $30\deg.8$ 
by black dots and their deconvolution into Gaussian components shown by red lines. The fitted center velocity of the outermost component is defined as the terminal velocity. } 
\label{fig2}
\end{figure}

The simplest way to determine the terminal velocity is to pick up the highest-velocity component after deconvolution of the line profile into many components.
Figure \ref{fig2} shows CO line spectra in the Galactic plane at several longitudes around $l=31\deg$.
Each spectrum can be expressed by superposition of many components, each represented by a Gaussian profile, as indicated by the red lines.  
The highest-velocity component is uniquely determined in each profile as the rightmost Gaussian component.
We then define the terminal velocity as the center value of the highest-velocity Gaussian component.

Before applying the deconvolution, we smoothed the data cube in the latitude direction by a Gaussian beam of 
$\delta l \times \delta b=25''\times 60''$
in order to increase the signal-to-noise ratio compared to the original data at $20''\times 20''$ resolution without much loosing longitudinal and velocity resolutions. 

\subsection{LV plot of terminal velocities}
\label{lvplot}

\begin{figure}  
    \begin{center}
    \includegraphics[width=8cm]{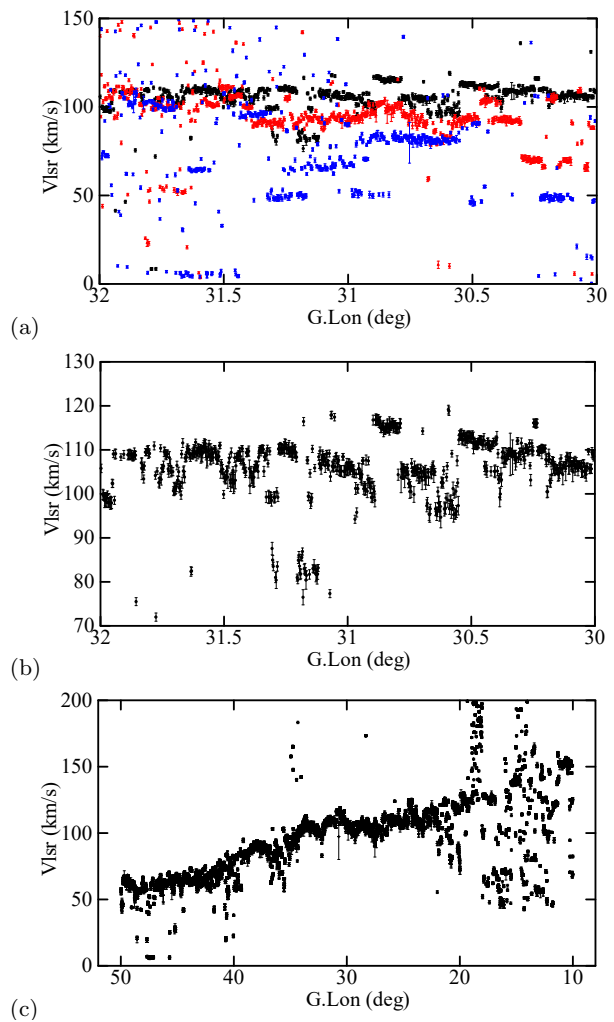} 
\end{center}
\vs \caption{(a) Terminal velocities at $b=0\deg$ (black dots), $0\deg.5$ (red triangles), and $0\deg.75$ (blue cross) from $l=30\deg$ to $32\deg$.
(b) Same, but for $b=0\deg$ with the vertical axis enlarged. (c) Terminal velocities from $l=10\deg$ to $50\deg$ at $b=0\deg$.} 
\label{fig3} 
\end{figure}

We apply the GDM to each spectrum of the \cotw line emission  of the CO data cube of FUGIN survey.
Figure \ref{fig3}(a) shows LV plots,  where the terminal velocities are plotted against longitude for the longitude range $30\deg$ to $32\deg$.  
The bars are the Gaussian-fitted full line widths of the components, which represent intrinsic velocity dispersions of clouds on the order of 5 to 10 \kms observed at the velocity resolution of 1.3 \kms.
The LV plots show that the terminal velocities  systematically decreases with latitude such that they are highest at $b=0\deg$ (red) and lowest at $b=0\deg.75$ in the plot in figure \ref{fig3}. 
In order to trace maximum terminal velocities in the disk, we below analyze the spectra in the galactic plane at $b=0\deg$. 

The latitudinal variation of terminal velocity may be attributed to the brightness decreasing with latitude: the higher becomes latitude, the higher becomes the possibility to detect fore/back-ground emissions as the terminal components due to the decreasing brightness.
Alternatively, the decreasing velocity with height from the Galactic plane is real, and represents slower rotation in the upper layer of the molecular disk.
 
 Figure \ref{fig3}(b) shows the LV plot enlarged in the velocity axis, revealing bumpy variation of the terminal velocity against longitude, with neighboring clumps often exhibiting discrete velocity jumps by 5 to 10 \kms.  
Such a variation represents proper velocity dispersion among the detected terminal-velocity clouds. 

    \begin{figure}   
    \begin{center}
    \includegraphics[width=8cm]{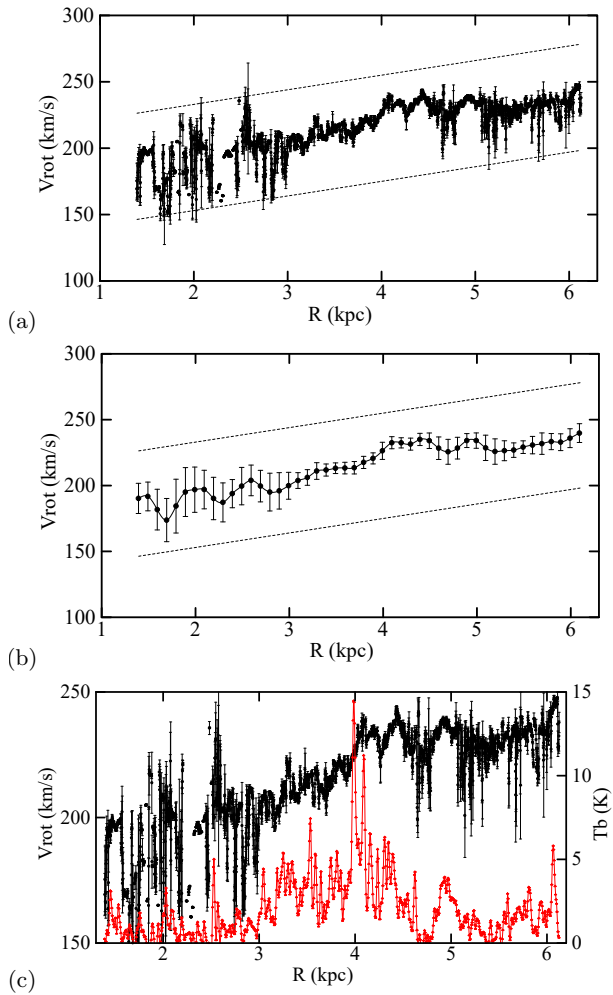}
\end{center} 
\caption{(a) CO-line RC from $R=1.3$ to 6.2 kpc after Gaussian running average with width of 2 pc.
Scattered data outside the two dashed lines ($\pm 40$ \kms from the mean) have been removed from the averaging.
(b) Same, but with 100 pc width.
(c) Same as (a), but enlarged and compared with $\Tb (\propto n_{\rm H_2})$ of the \cotw line.  
}
\label{fig4} 
\end{figure}

Figure \ref{fig3}(c) shows the thus obtained LV plot from $l=10\deg$ to $50 \deg$, where the terminal velocities are well determined at $l>\sim 20\deg$.
Larger scatter at $l<\sim 20\deg$ is not only because of the less sensitive observations, but also due to the intrinsically weak CO-line emission in the innermost region.

\subsection{FUGIN CO-line RC}

Since the terminal-velocities read from the data at higher latitudes tend to lead to lower-velocity rotation curve, we here measure the terminal velocities at $b\sim 0\deg$.
Using the LV plot in figure \ref{fig3}(c), we thus obtain a plot of maximum terminal velocities against longitude, which we adopt as the circular velocities.
Figures \ref{fig4}(a) and (b) show Gaussian running-averaged plots of the circular velocities, or the rotation curves, where both the radius interval and Gaussian half width were taken to be 2 pc and 100 pc, respectively.
In figure \ref{fig4}(c) we enlarge the velocity axis, and compare the RC with CO-line $\Tb$, which is proportional to the local molecular gas density.

During the averaging, we removed data points exceeding $\pm 40$ \kms from the expected mean value as approximated by a linear function,
 $V_{\rm mean}({\rm km ~ s^{-1}})= 225 + 11(R({\rm kpc})-4)$ \kms,  
as indicated by the dashed lines in order to avoid the anomalously deviated data.

\section{Discussion}

\subsection{Comparison with the current RCs} 

 The FUGIN RC coincides well with the current RCs \citep{Sofue2017,sofue2020} (see the referenced papers therein for RC compilation from the literature)  within a few \kms at radii greater than 3.5 kpc.
Detailed behavior with bumpy fluctuations is found to be very similar to that of the inner RC of the 4th quadrant from HI tangent-velocity measurements \citep{mcclure+2007}. 

However, the FUGIN RC tend to show systematically lower values than the current RCs at radii less than 3.5 kpc.
The same trend is found against the southern RCs in HI and CO lines observed in the 4th quadrant  \citep{alvarez+1990,mcclure+2007}. 
This may be attributed either to real difference from the current RCs obtained at different resolutions and/or in the different Galactic quadrant, or due to an artifact caused by larger scatter at $l < \sim 20\deg$, where we had less accurate fitting because of the weaker or almost vacant CO emission.  
 
\subsection{Local RC variations}
 
The RC is superposed by wavy and bumpy fluctuation around the mean with amplitude at $\delta V\sim 10$ \kms. 
The largest bump is found with its peak at $R=4.1$ kpc, associated with the Scutum arm.
The velocity increases steeply from 210 to 230 \kms
between 3.9 and 4.1 kpc, followed by a plateau-like enhancement till $\sim 5.2$ kpc. 
The velocity plateau is superposed by two more bumps at $R=4.4$ and 4.9 kpc. 
Also at $R=3.1$ kpc there is a velocity depression followed by a jump to the peak at 3.3 kpc.   
These velocity bumps are similar to those found in the 4th quadrant both by their amplitudes and scale lengths \citep{mcclure+2007}. 

Besides such prominent bumps, there appear superposed smaller fluctuations of radial scale length of $\delta R \le \sim 0.1$ pc and velocity amplitudes of $\delta V\sim \pm 2-3$ \kms.
These small bumps may be attributed to random motions of molecular clouds near the tangent points.
As already discussed the velocity fluctuations found inside 2.6 kpc can be attributed to noisy data because of the weak CO-line emission.

\subsection{Imprint by arms}

The RC fluctuations with wave lengths of $\sim 1-2$ kpc can be attributed to an imprint by spiral arms associated with non-circular streaming motion \citep{roberts1969,martinez+2020}, or to local gaseous ring and vacancy \citep{Sofue+2009,mcgaugh2019}. 
In either mechanism, the conservation of angular momentum  results in deceleration of rotation velocity for the gas flowing from the inside to the arm, and acceleration for the gas from outside.

Let $\delta R$ and $\delta V$ be the radius and velocity displacements from the purely circular motion at $R$ with $\Vrot$.
Neglecting the second order quantities, we have 
$  \delta V/\Vrot \simeq - \delta R/R$.
In order to attain the observed velocity acceleration by $\sim \pm 9$ \kms for $\Vrot=220$ \kms, the gas must be accumulated from a region in $\sim \mp 0.36$ kpc around the density peak in the arm.
Such gas flows from both sides of the arm center cause a steep velocity gradient as observed around $R\sim 4$ kpc in figure \ref{fig4} to yield $dV/dR \sim 70$ \kms kpc$^{-1}$.
This is observed as the positive gradient of LV ridge around W43 in the tangential direction of the 4-kpc (Scutum) arm with $d \vlsr/dl \sim 10$ \kms per degree in figure \ref{fig1}(c), where the expected gradient of the terminal velocity for flat rotation is negative with $\sim -3.6$ \kms per degree.  

In figure \ref{fig4}(c) the brightness temperature $\Tb$ of the \cotw line at tangential velocities is plotted by the red line.
The 4-kpc arm shows up as the sharpest $\Tb$ peak at $R=4.0$ kpc, nesting the molecular complex and star forming region W43 \citep{kohno+2021}.
Slightly inside this peak at $R\sim 3.9$ kpc, $\Vrot$ is observed to attain a local minimum about $\delta V \sim -10$ \kms displaced from the mean, indicating deceleration of the gas due to accumulation toward the arm center at 4-kpc density peak.
An opposite-sense behavior with $\delta V\sim +10$ \kms at 4.1 kpc is observed outside the peak.

\subsection{Galactic shock wave in the Scutum Arm}  
\label{spiralarm}
 
\begin{figure}  
   \begin{center}
   \includegraphics[width=8.5cm]{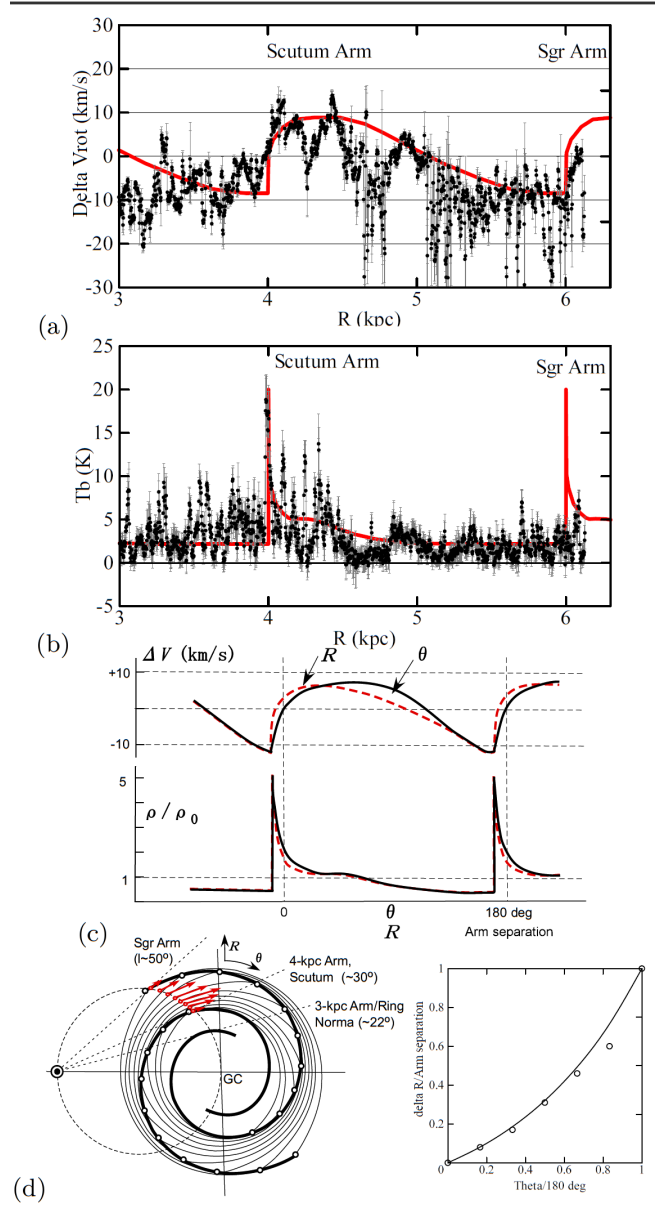} 
\end{center}   
\caption{(a) Observed (dot with error bar) and theoretical (red line) $\Delta V$  against radius $R$.
(b) Same, but for 
$\Tb (\propto \rho)$.
(c) GS model of $\delta V$ and $\rho$   \citep{roberts1969} against $\theta$ (solid) and $R$ (dash, schematic).  
(d) Spiral shocked arms (thick lines) and flow lines (thin lines) representing the Scutum and Sgr Arms. The right panel shows relation between $\delta R$ and $\theta$ in the GS model.
}  
\label{fig5}
\end{figure}

Figure \ref{fig5}(a) shows variation of the RC excess from the mean, 
$\Delta V=\Vrot -V_{\rm mean}$,
where the mean around Scutum and Sgr Arms was approximated by $V_{\rm mean}=225+10~(R({\rm kpc})-4)$ \kms.
The velocity amplitude is measured to be $\sim \pm 9$ \kms. 
Panel (b) shows the CO brightness temperature $\Tb~ (\propto n_{\rm H_2})$, which is proportional to the local volume density of the molecular gas.
The density compression at 4 kpc, which is supposed to be the shock front, is measured to be $\rho/\rho_0\sim \Tb /T_{\rm B,mean} \sim 20~{\rm K}/4~{\rm K}\sim 5$, and the width $\delta R \sim 0.1$ kpc. 

Panel (c) shows theoretical curves of $\Delta V$ and  $\rho/\rho_0$ plotted against azimuth angle $\theta$ along a flow line as reproduced from Roberts (1969).
The red dashed lines schematically represent those against the radius. 
Here, the radial distance $\delta R$ is related to azimuth angle $\theta$ using the plot of a flow line through the spiral arms presented by Roberts (1969).
The relation is, here, empirically expressed by 
$\delta R \sim \Delta (e^{3x/2}-1)/(e^{3/2}-1)$,
where $x=\theta/\pi$, and $\Delta$ is the separation between the arms as explained by figure \ref{fig5}(d).
Note that the radial variation is much sharper than azimuthal variation. 
 
The red lines in figures \ref{fig5}(a) and (b) are calculated radial profiles of the velocity and density based on the galactic shock wave theory, where the shock front is put at $R=4$ kpc and the arm separation is taken to be $\Delta=2$ kpc corresponding to Scutum and Sgr arms.  Here, we recalled the well known three major arms in the 1st quadrant, the Norma (3-kpc, $l\sim 20\deg$), Scutum (4-kpc, $\sim 30\deg$), and Sgr ($\sim 50\deg$) Arms with tangential radii $R\simeq 3$, 4, and 6 kpc, respectively \citep{nak+2016}.

 We emphasize that the observed RC and density properties of the Scutum Arm are well fitted by the GS model: namely, the velocity bump at 4 kpc followed by a plateau, velocity amplitude $\pm 9$ \kms, sharp and narrow density peak at 4 kpc, and density compression $\rho/\rho_0\sim 10$. 
The Scutum arm is, therefore, naturally explained by an ordinary galactic shock wave in a normal spiral arm with potential depth corresponding to $\sim \pm 9$ \kms. 

We comment that this simple view of spiral arm does not contradict the anticipated bar potential and induced kinematics \citep{wein1992,binney+1991,athana+1999}, if the bar end is located sufficiently inside $R\sim 4$kpc and the arm is stretched outside the bar end at small pitch angle.

\section{Summary}
  
We analyzed the FUGIN CO-line data to determine the rotation curve at the highest resolution ever obtained in the inner Milky Way. 
The RC is found to be superposed by small-scale variations with velocity gradients as high as several tens \kms kpc$^{-1}$ and amplitude $\pm 9$ \kms. 
The local RC and density variations toward the 4-kpc arm is understood as due to a galactic shock wave in a normal spiral arm, not requiring a strong shock by a bar potential.  
  
\vskip 2mm
\noindent {\bf Acknowledgements/data availability}:
Data analysis was carried out at the Astronomy Data Center of the NAOJ. 
CO data were taken from the FUGIN survey with the Nobeyama 45-m telescope available at the URL: http://nro-fugin.github.io. 
Tables of RCs are available as a PASJ supplementary data, and at 
http://www.ioa.s.u-tokyo.ac.jp/$\sim$sofue/h-rot.htm.

\end{document}